# Developments in International Masterclasses

K. Cecire
*University of Notre Dame, Notre Dame, IN 46556, USA*

International Masterclasses are events for high school students that enable them to act as "particle physicists for day" by analyzing authentic data from particle physics experiments under the tutelage of scientists. International Masterclasses have been very successful and have grown with their focus on data from the Large Hadron Collider at CERN. Recent developments that bring masterclasses to more students in more ways have been very exciting. The number of masterclasses and the number of countries with masterclasses have grown substantially. Outreach efforts such as masterclass workshops in developing countries, particularly in collaboration with the African School of Fundamental Physics and Applications, and a new International Day of Girls and Women in Science masterclass have yielded success. Most recently, there are new initiatives to expand masterclasses beyond the LHC, particularly into neutrino physics.

## 1. INTRODUCTION

### 1.1. Understanding International Masterclasses

The term "International Masterclasses" (IMC) refers to both the masterclasses in particle physics that are held at universities and laboratories around the world within a set interval of time each year (usually about five to seven weeks centered in March) and the International Masterclasses organizational structure that coordinates, schedules, and supports these masterclasses.

A masterclass at a university or laboratory is a daylong experience for high school students. In this day, students attend presentations on experimental particle physics and how to analyze data. These presentations focus the students on measurements from a particular experiment that they will analyze. Currently, those experiments are from the four main detectors in the Large Hadron Collider (LHC) at CERN: ALICE, ATLAS, CMS, and LHCb. Students also take tours of physics facilities within the institution they are visiting. The main event is the measurement itself. Students examine data in the form of event displays, event-by-event, categorizing the events and building up statistics by combining their results. For example, in the ATLAS Z measurement, students work in pairs to search for dilepton signals and pick out the two leptons in each event that are most likely to come from a single neutral parent. Their software generates an invariant mass from each pair of leptons chosen and then a text file of these masses with the flavors of the leptons. Students upload this file to an online platform that takes in the data for all of the students doing this same analysis on that day and generates mass plots. In recent years, this measurement has been extended to include a search for Higgs-like signals. [1]

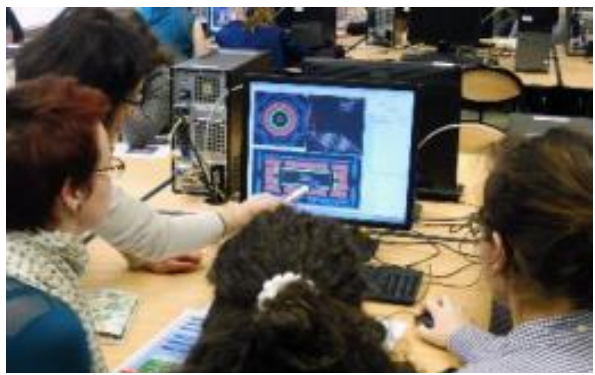

Figure 1: Students examine an ATLAS event display.

An important key to the measurement phase is the active participation of particle physicists as tutors to the students, helping them with the mechanics of making the measurement and, most importantly, advising them on interpretation of events. Students learn a great deal about the experiment and about how physics is done when they ask about events that do not seem to fit the pattern they were taught. Tutors must rely on their own experience as well as understanding the detector to help students and, in so doing, impart some of that understanding to students. Sometimes two tutors will disagree on the interpretation of an event. The best tutors lead students to work out their own reasonable answer. Students see that nature does not provide an "answer key" and learn something of how physicists look at data.

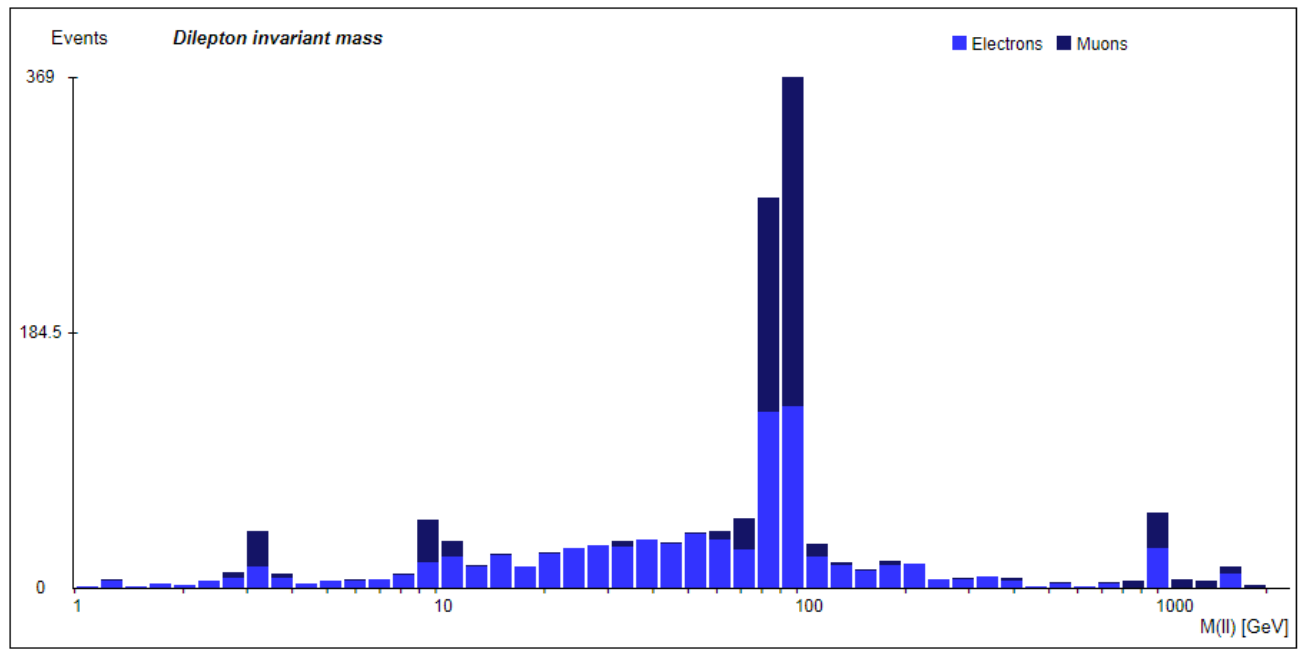

Figure 2: Dilepton plot for ATLAS Z masterclass measurement by four masterclass institutions on April 7, 2017. Please note signals for well-known mesons at lower masses and Monte Carlo "discoveries" at very high masses.

At the end of the masterclass day, students at a masterclass institution have a discussion with the tutors to draw conclusions and understand the physics behind their results. In the previously mentioned ATLAS Z measurement, students will see several mass plots that help them find the mass of the Z boson but also discover other particles in the data which also decay to dileptons. After this rich physics discussion, the students engage in a videoconference with up to four other masterclass institutions which made the same analysis (but with different datasets) and physicist moderators at CERN, Fermilab, or, for one day per year, TRIUMF. Over the course of an "IMC season," there are generally around 100 such videoconferences.

### 1.2. International Masterclasses in 2017

International Masterclasses have become one of the most important components of worldwide particle physics outreach and education. In 2017, International Masterclasses ran from March 1 through April 11. In that time, there were 314 masterclasses at 116 institutions for over 10,000 students in 50 countries. [2] In addition, there were masterclasses throughout the year that were not in the IMC count and may or may not have had videoconferences. While the language of most videoconferences is English, the masterclasses are done on the local level in many languages. The ATLAS and CMS masterclass websites are each offered in thirteen languages; ALICE is in seven languages and LHCb in six languages. [3]

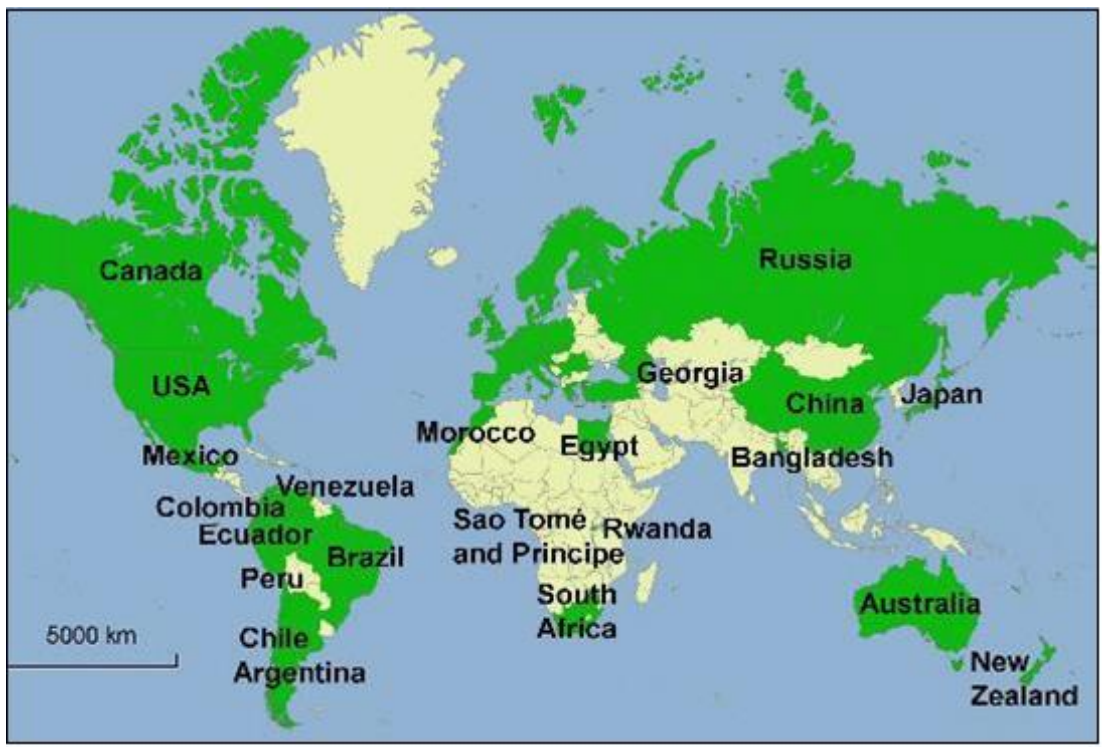

Figure 3: Map showing countries participating in IMC 2017 in green.

## 2. LHC MEASUREMENTS

There are currently masterclass measurements for four LHC experiments.

The ALICE experiment has two measurements. The most popular is the ALICE Strange Particles measurement. Students look for weak decays of strange particles (that is, hadrons with at least one strange quark) and make mass plots to identify which decays indicate strangeness. Students then establish a ratio of strange particles to "normal" particles made of up and down quarks in the data sample. An enhancement of strangeness in the sample is an effect of the high temperature in the quark-gluon plasma that is studied in ALICE. [4] The second measurement, used less often but still of great interest, is the Nuclear Modification Factor $R_{AA}$ measurement. Students determine $R_{AA}$, the ratio of particles produced in lead-lead collisions to those produced in proton-proton collisions. [5]

ATLAS also has two measurements. The ATLAS Z measurement is described above. The ATLAS W measurement looks for two effects: the ratio of $W^+$ to $W^-$ bosons produced in proton-proton collisions and the prevalence of $W^+W^-$ events at different opening angles. With sufficient statistics, it may be possible to see evidence of the Higgs in any excess over the expected background, though in practice this is very difficult to achieve in the masterclass. The $W^+$ to $W^-$ ratio is a good probe of both proton structure and the behavior of the detector. At the simplest level, protons have more positively charged quarks than negatively charged quarks; the W bosons are made in collisions of these quarks, so the ratio should be greater than one. Looking at it with more detail, $W^+$ bosons can come from the collisions of up valence quarks with anti-down sea quarks or from up and anti-down sea quarks. $W^-$ bosons can come from collisions of down valence quarks and anti-up sea quarks or down and anti-up sea quarks. Since there are twice as many up quarks to down quarks at the valence level of the proton, the maximum $W^+$ to $W^-$ ratio should be 2. Since up, anti-up, down, and anti-down sea quarks all appear with the same probability, the minimum $W^+$ to $W^-$ ratio should be 1. The details of exactly where this ratio falls is due to further details of proton structure and detector effects. [6]

The CMS experiment also has two measurements. The first, the J/Ψ measurement, was used in the first year of LHC masterclasses (2010) when J/Ψ decays into dimuons were the only CMS events available to students and the public. In this measurement, students examine event displays to make "quality cuts" on the data; events are rated by the likelihood that the muon pairs come from a single neutral particle like the J/Ψ meson. Events with two like-sign muons are eliminated outright from consideration. Others are rated 1 to 3 depending on the quality of the tracks. (Events rated "3" have the highest quality.) Students can compare mass plots with looser (2 and 3 allowed) or tighter (3 only) cuts. This measurement is still used in some school situations and independent masterclasses. The second, the WZH measurement, is similar to the ATLAS Z measurement but with some important differences. There are no Monte Carlo "discoveries"—all of the data is authentic—and students create a mass plot as well as find the $W^+$ to $W^-$ ratio. Also, there a few Higgs candidates mixed in for students to find and even include in the results. Operationally, the CMS WZH measurement can be done entirely online. The event display, iSpy-WebGL, and the analysis tool, CMS Instrument for Masterclass Analysis (CIMA), are both accessible via the Internet by means of a browser. In CIMA, students click the invariant masses into the appropriate 2 GeV bins to build the mass plot themselves; the mass plot is not built automatically. [7]

The CMS WZH measurement was upgraded in 2017. The invariant mass in previous versions was given automatically in CIMA when an event was classified dielectron, dimuon, diphoton, of 4-lepton by students. The newer version retains this for Higgs candidates but for dileptons, students must now select the leptons in iSpy-WebGL. When they do this, an invariant mass appears, which they must enter into CIMA. As an alternative for a teacher using iSpy-WebGL, students can choose lepton tracks and get 4-vector information so that it is possible to calculate the invariant mass by hand or in a spreadsheet as an exercise. The number of events was also increased so that there is much less oversampling of data within masterclasses. (The ATLAS Z measurement does not have this problem due to the already large number of available events.) In 2018, Drell-Yan background will be incorporated into the data so that students can see its structure. In 2019, the QuarkNet CMS masterclass team hopes to introduce a Higgs masterclass using diphoton, 4-lepton, $W^+W^-$, and bottom-antibottom events.

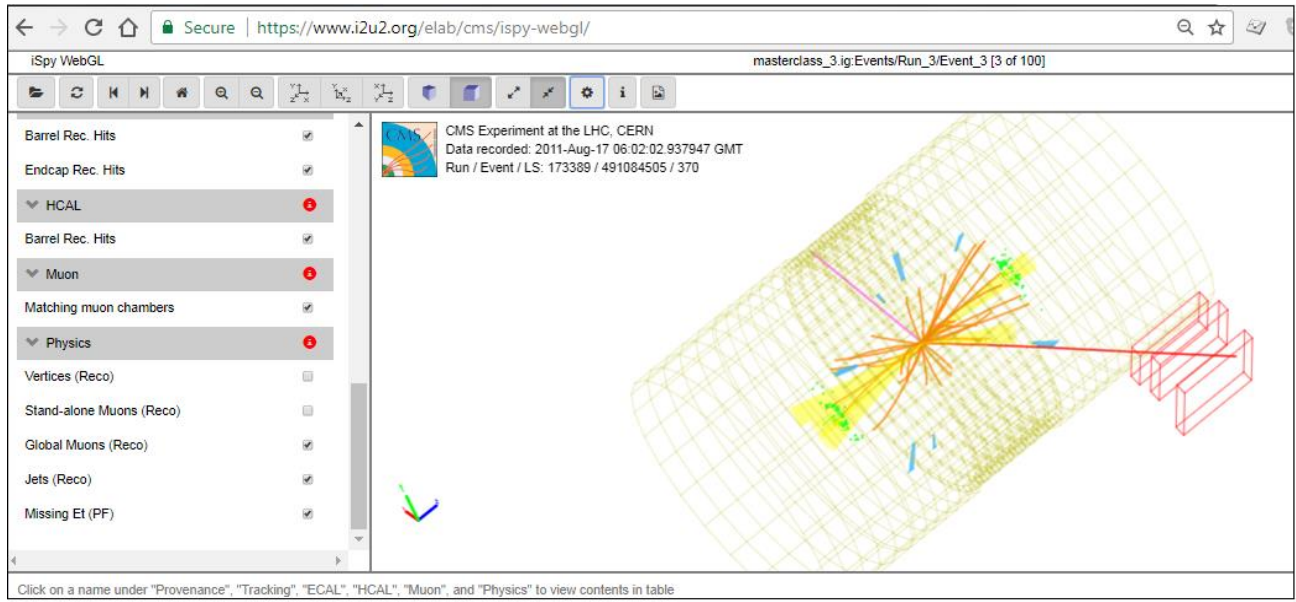

Figure 4: CMS event display iSpy-WebGL in inverted colors showing background tracks, missing $E_t$, and a muon track.

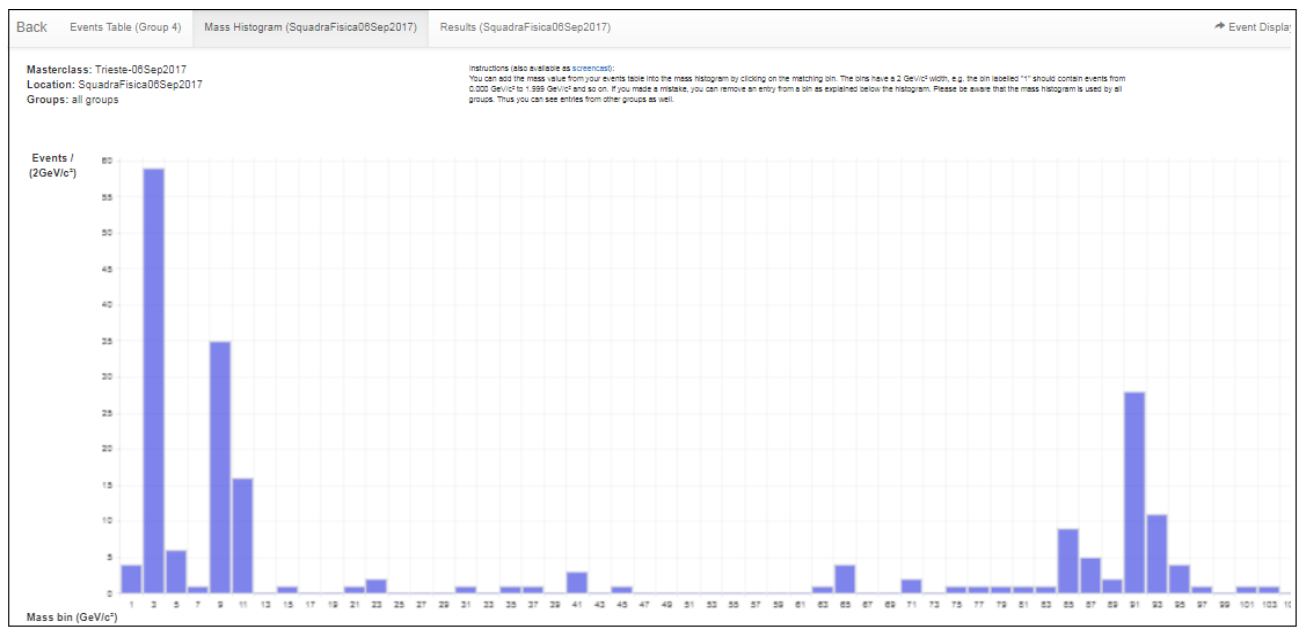

Figure 5: CMS WZH measurement mass plot from CIMA.

The LHCb experiment has one measurement in which students look for and measure displaced vertices in decays of mesons with B quarks. Students use the energy and momentum of the $D_s$ meson and the length of its displaced vertex— the distance it travels before it decays —to calculate its lifetime. [8]

All International Masterclasses have some features in common. They are one-day events, though teachers who bring students to masterclasses often give them some preparation ahead of time (QuarkNet recommends three hours; its internal data shows this improves student understanding and satisfaction), and students work in pairs—two to a computer and data set—to check each other as they work as well as make the analysis more efficient. The interaction of physicists and students is an important linchpin and is where most of the learning takes place. Finally, the videoconference is the "fun" at the end but also cements the idea that particle physics works because of large, international collaborations.

## 3. EXPANDING MASTERCLASSES

### 3.1. Neutrino Physics in IMC

Members of QuarkNet, the International Particle Physics Outreach Group (IPPOG), and several notable accelerator-based neutrino experiments (notably MINERvA, MicroBooNE, and DUNE) have contributed to an emerging plan to introduce neutrino masterclasses to IMC.

It should be noted that a non-accelerator experiment, IceCube, already has a masterclass that it operates with a growing number of institutions and students. [9] The IceCube masterclass effort is led by Silvia Bravo-Gallart, James Madsen, and collaborators from WIPAC, the Wisconsin IceCube Particle Astrophysics Center at the University of Wisconsin. The IceCube masterclass is not affiliated with IMC, but the two groups maintain ongoing contact directly and through QuarkNet.

The development of an accelerator-based neutrino masterclass begins with the MINERνA experiment. MINERνA has a measurement for student data analysis in place thanks to the efforts of Kevin McFarland and his collaborators at the University of Rochester. [10] This same team is currently building a larger, more varied dataset to make a robust masterclass measurement. Using this new dataset, students will separate signal from noise and then use momentum conservation from weak interactions of muon neutrinos with neutrons to probe the momenta of neutrons inside the nucleus of the carbon target. This is then used as an indirect probe of the size of the nucleus. It is expected that MINERνA masterclasses will be introduced as part IMC 2018.

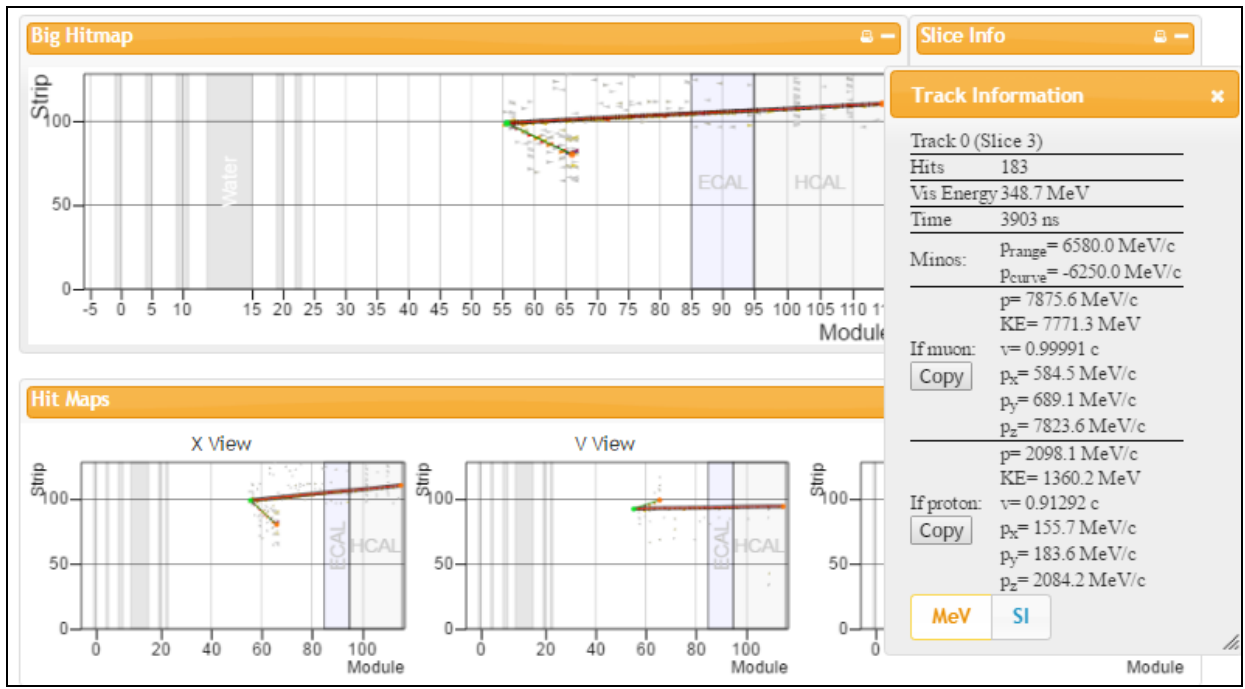

Figure 6. MINERνA event in the Arachne event display.

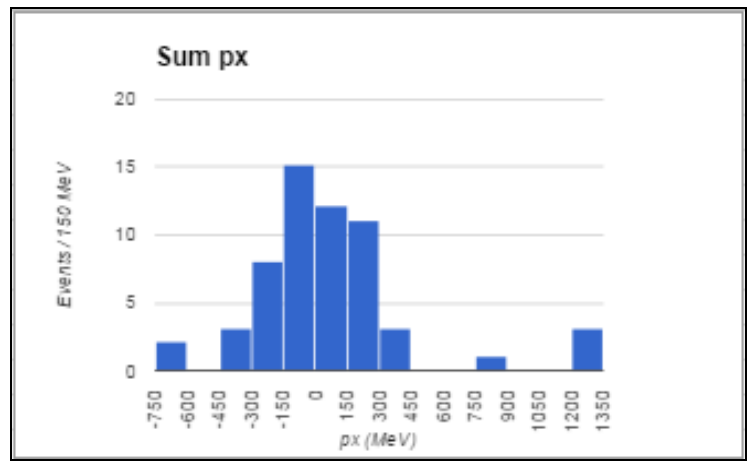

Developments in International Masterclasses

Figure 7. Combined result of the sort to be used in the MINERvA masterclass, piloted by QuarkNet teachers in August 2016. Students use the standard deviation of the momentum distribution and the Heisenberg Uncertainty Principle to estimate the distance distribution of neutrons in the carbon nucleus, which in turn is an estimate of the diameter of the nucleus.

Sowjanya Gollapinni of the University of Tennessee Knoxville leads the development of a MicroBooNE masterclass in which students will use authentic data to measure argon purity in the detector and then make multiplicity measurement from interactions of neutrinos with nuclei. We expect a test version of this measurement in 2018 for inclusion in IMC 2019.

Beyond 2019, the goal is to use authentic data from DUNE when it becomes available. With DUNE masterclasses, students will be able to tap into rich physics and engineering context. Student measurements will include neutrino oscillations along with other rich physics and engineering questions.

It is important to note that there are other neutrino experiments that might still contribute to the developing neutrino masterclasses effort. The experiments mentioned in this paper are already at work on measurements and have been part of the planning process. Others are welcome to join.

### 3.2. Outreach and Collaboration

IMC has a history of reaching out to new student and teacher audiences. This initially took the form of expansion to more institutions in countries with ties to LHC experiments. As this continues, new initiatives are reaching groups that have not previously had access to masterclasses.

IMC piloted World Wide Data Day (W2D2) in 2016 as a way to give students a short, simple masterclass experience in school with authentic LHC data. It takes place during a single day, midnight-to-midnight UTC, with students analyzing data under direction of their teachers and connecting to short videoconferences with physicist moderators taking "shifts" according to their time zones. Students look for dimuon events in CMS and ATLAS masterclass data— these are readily discoverable with no adjustments to the event display—and measure muon direction angles $\phi$ and $\theta$. The students contribute to histograms of the direction angles that they build over the course of the day; students and teachers discuss the meanings of these simple but significant plots with physicist moderators in the videoconferences. The initial W2D2 pilot took place on December 2, 2016 with approximately 350 students at 20 schools across the globe. In 2017, W2D2 will be held on November 14 with, it is hoped, 100 groups. [11]

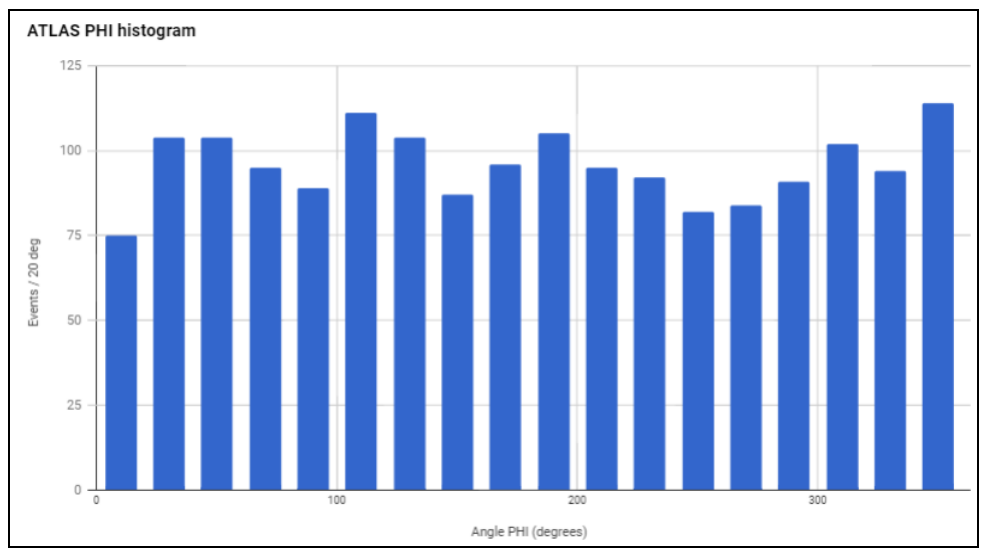

Figure 8. W2D2-2016 combined result for distribution of muon tracks in ϕ from the ATLAS detector.

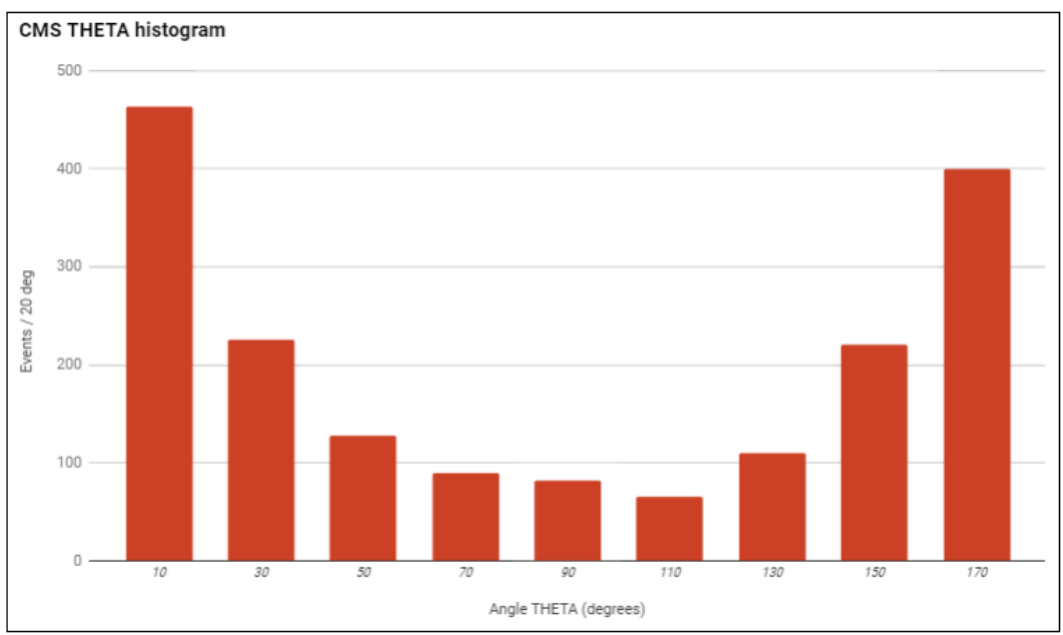

Figure 9. W2D2-2016 combined result for distribution of muon tracks in θ from the CMS detector.

On February 10 and 11, 2017, IMC also piloted special masterclasses for the International Day of Women and Girls in Science (IDWGS). These were masterclasses from all four main LHC experiments for approximately 320 girls in 10

Developments in International Masterclasses

masterclasses. All of the students were girls and, to the extent possible, the physicists involved were women. Videoconference discussions included topics relating to studying science and being a scientist as a woman. These were received with great enthusiasm; this program will continue with special masterclasses on February 12, 2018 and is expected to grow. [12]

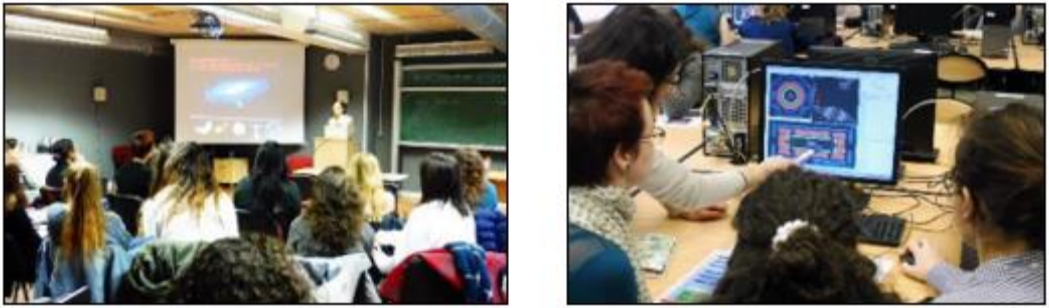

Figure 10. Scenes from IDWGS 2017 masterclasses.

IMC has begun to collaborate with the African School of Fundamental Physics and Applications (ASP) to provide a program of outreach and education which also reaches new groups with masterclasses. When ASP was held in Kigali, Rwanda in August 2016, associates of IMC came to deliver a day-and-a-half workshop for teachers and three short workshops for high school students. The teacher workshop included a CMS masterclass measurement as well as short ATLAS and CMS masterclass measurements for schools with little or no technology support that teachers can bring to their students. The student workshops also included these shorter masterclass measurements. Thus, approximately 150 students and 20 teachers in Rwanda were exposed to masterclasses and had a chance to work with LHC data. At the same time, Kate Shaw, representing Physics without Frontiers, ICTP, and IMC, ran an ATLAS masterclass for the graduate students attending ASP 2016. In addition, part of the IMC team then traveled to Addis Ababa, Ethiopia and did another CMS masterclass there. [13]

ASP and IMC are already planning an outreach and education program for ASP 2018 in Windhoek, Namibia. Members of the Namibian National Commission on Research Science and Technology (NCRST) were in attendance at ASP 2016. Namibia has asked for an extended program: a five-day teacher workshop and four to eight high school student workshops.

IMC has a history of collaboration to expand the reach of masterclasses. There has been an Expanding Masterclasses Working Group as part of IPPOG since 2015 to plan and coordinate some of these efforts, but the collaborations run deeper. In 2014, ATLAS, CMS, and the University of Notre Dame contributed to funding Masterclass Institutes Collaborating in the Americas, which took masterclass workshops for teachers to Chile, Colombia, Ecuador, Jamaica, and Mexico. All of these, aside from Jamaica, continue to be active in International Masterclasses. Local mentors and Physics without Frontiers were able to work with IMC to bring additional masterclasses to Venezuela, Peru, and Argentina. Notre Dame has continued funding the effort with Chile and, more recently, Mexico; there will be further masterclass workshops in new places in these two countries in 2018. [14]

## 4. ORGANIZATION

Developments in International Masterclasses

Several groups are responsible for International Masterclasses. IMC is supported by IPPOG, CERN, Fermilab with coordination by Uta Bilow at the Institut für Kern- und Teilchenphysik at Technische Universität Dresden and Kenneth Cecire at the University of Notre Dame QuarkNet Center. They, in turn, are supported by Netzwerk Teilchenwelt in Germany and QuarkNet in the United States. Both groups have had key roles in developing and promoting masterclasses.

## Acknowledgments

The author wishes to thank Uta Bilow, Technische Universität Dresden, for key content and LaMargo Gill of Fermilab for editing.

This work is supported by QuarkNet and the National Science Foundation.